\documentclass[conference]{IEEEtran}
\usepackage{blindtext, graphicx}
\usepackage{booktabs,multirow,array}
\newcolumntype{N}{@{}m{0pt}@{}}
\usepackage{fancyhdr}

\usepackage{amsmath}
\usepackage{amssymb}
\usepackage{graphicx}
\usepackage{epstopdf}
\usepackage{cite}
\usepackage[ruled, vlined]{algorithm2e}
\usepackage{url}
\usepackage{tikz}
\usetikzlibrary{patterns}
\usetikzlibrary{shapes.misc}
\usepackage{multirow}
\usepackage{xcolor}
\usepackage{subcaption}
\usepackage{accents}

\graphicspath{ {./images/} }

\usepackage{lastpage}
\usepackage{adjustbox}
\usepackage{array}

\newcolumntype{R}[2]{%
    >{\adjustbox{angle=#1,lap=\width-(#2)}\bgroup}%
    l%
    <{\egroup}%
}
 
\begin{document}
%
\title{Initializing Successive Linear Programming Solver for ACOPF using Machine Learning}
\author{Sayed Abdullah Sadat, Graduate Student Member, IEEE, Mostafa Sahraei-Ardakani, member, IEEE, \\
Department of Electrical and Computer Engineering, University of Utah  
\\
email: \textit{sayed\_abdullah@ieee.org}} 



\maketitle

\begin{abstract}

A Successive linear programming (SLP) approach is one of the favorable approaches for solving large scale nonlinear optimization problems. Solving an alternating current optimal power flow (ACOPF) problem is no exception, particularly considering the large real-world transmission networks across the country. It is, however, essential to improve the computational performance of the SLP algorithm. One way to achieve this goal is through the efficient initialization of the algorithm with a near-optimal solution. This paper examines various machine learning (ML) algorithms available in the Scikit-Learn library to initialize an SLP-ACOPF solver, including examining linear and nonlinear ML algorithms. We evaluate the quality of each of these machine learning algorithms for predicting variables needed for a power flow solution. The solution is then used as an initialization for an SLP-ACOPF algorithm. The approach is tested on a congested and non-congested 3 bus systems. The results obtained from the best-performed ML algorithm in this work are compared with the results of a DCOPF solution for the initialization of an SLP-ACOPF solver.   
\end{abstract}


%
\IEEEpeerreviewmaketitle
\section{Introduction}
{\let\thefootnote\relax\footnote{{This research was funded by the U.S. Department of Energy's (DOE) Advanced Research Projects Agency--Energy (ARPA-E) under grant number DE-AR0001083.}}}
The primary goal of power system operation is to serve the customers at the minimum cost, taking into consideration the physical constraints of the system. This can be expressed as an optimization problem, referred to as the optimal power flow (OPF). The original ACOPF problem is a nonlinear and non-convex optimization problem, which is an NP-hard problem \cite{castillo2013computational,cain2012history,sadat1}. The challenges include the difficulty to identify a globally-optimal solution, specifically its computational burden for large scale real world networks. The computational expense becomes manifold when we consider the North American Electric Reliability Corporation's (NERC) N-1 reliability criteria and account for contingencies \cite{sadat2,sadat2018reducing}. As a result, different linearization techniques are used by system operators, most commonly the Direct Current Optimal Power Flow (DCOPF) approach \cite{stott2009dc, sayed, sadat2017optimal}. Since DCOPF abstracts from some of the complexities of the original problem, it does not produce a physically-feasible solution. Thus, the operators need to adjust the DCOPF solution, in order to achieve feasibility \cite{al2015role}. The simplifications associated with the use of DCOPF along with out of market adjustments can result in large inefficiencies in the final solution. A staff report by the Federal Energy Regulatory Commission (FERC) estimates that the status quo may result in up to 10\% additional cost, which can be avoided through efficient ACOPF solvers \cite{cain2012history}.

Considering the size of the North American power grid, a small percentage of improvement in its operations can yield a significant monetary savings~\cite{seth}.  
That's why there is a growing interest among academic scholars to work around techniques that would solve the ACOPF fast enough for the real-time operations, and thereby operating the system more efficiently. The challenge remains to be the computational complexity of finding a quality solution within the limited available time. 

Many methods have been proposed and tested for solving ACOPF, including a variety of convex relaxation techniques \cite{lavaei2012zero,madani2015convex,low2014convex,hijazi2017convex}. Another approach that is shown to perform well is the successive-linear programming (SLP) approximation of the current‐voltage (IV) ACOPF formulation \cite{oneill}. This method takes advantage of the linear representation of network flows in a rectangular IV formulation, compared to a conventional quadratic  power flow formulation in polar coordinate. The SLP IV-ACOPF algorithm demonstrates promising scalability and performance properties~\cite{anya}. 

On the other hand, the recent advancements in computational capabilities have opened the door for many industries to effectively utilize the large quantities of data available. Power system operations and planning is no exception to this trend. The work in \cite{mlreview} provides a review of machine learning applications in energy machine learning systems. Learning optimal power flow is one area of the areas that has recently attracted interests in academia \cite{mlopf}.

In our previous work \cite{sadat1}, we have shown that the performance of an SLP solver, in terms of convergence quality, objective value, and computational performance, largely depends on the initialization of the SLP algorithm. Therefore, in this paper we would like to evaluate how various machine learning algorithms can perform as an initialization to consequently enhance the performance of the SLP method for solving an ACOPF problem. The approach can be considered as a special case of warm-start where the model utilizes the existing data for training and based on previous solutions, it predicts an operation point for the initialization of an SLP algorithm. 

This paper is aimed for the industry operation of power systems that takes advantage of the available data to improve the computational performance of an SLP algorithm for solving a large scale ACOPF problem. The contributions of this paper can be summarized as follows: 
\begin{enumerate}
\item Evaluation of different machine learning algorithms available in the Scikit-Learn library for an optimal power flow problem. 
\item Comparing the prediction by the best-performed algorithm with a DCOPF solution for a both congested and non-congested 3 bus systems.
\end{enumerate}

To obtain a feasible initialization, this paper proposes the use of machine learning algorithm and power flow to find a feasible and a near-optimal solution for initializing an SLP-ACOPF solver. Similarly, the results obtained by the DCOPF for the comparison is also solved by a power flow solver. The rest of the paper is organized as follows. In Section II, we present and introduce different popular regression algorithms available in the Scikit-Learn library. Section III presents the results of the evaluation of the quality of prediction by different machine algorithms. It also benchmarks the results with the solution of DCOPF for the initialization of an SLP solver. Section IV discuses the results and addresses some of the challenges associated with the use the proposed method, and finally conclusions are drawn in Section V.

\begin{figure}[h!]
\centering
\includegraphics[scale=0.15]{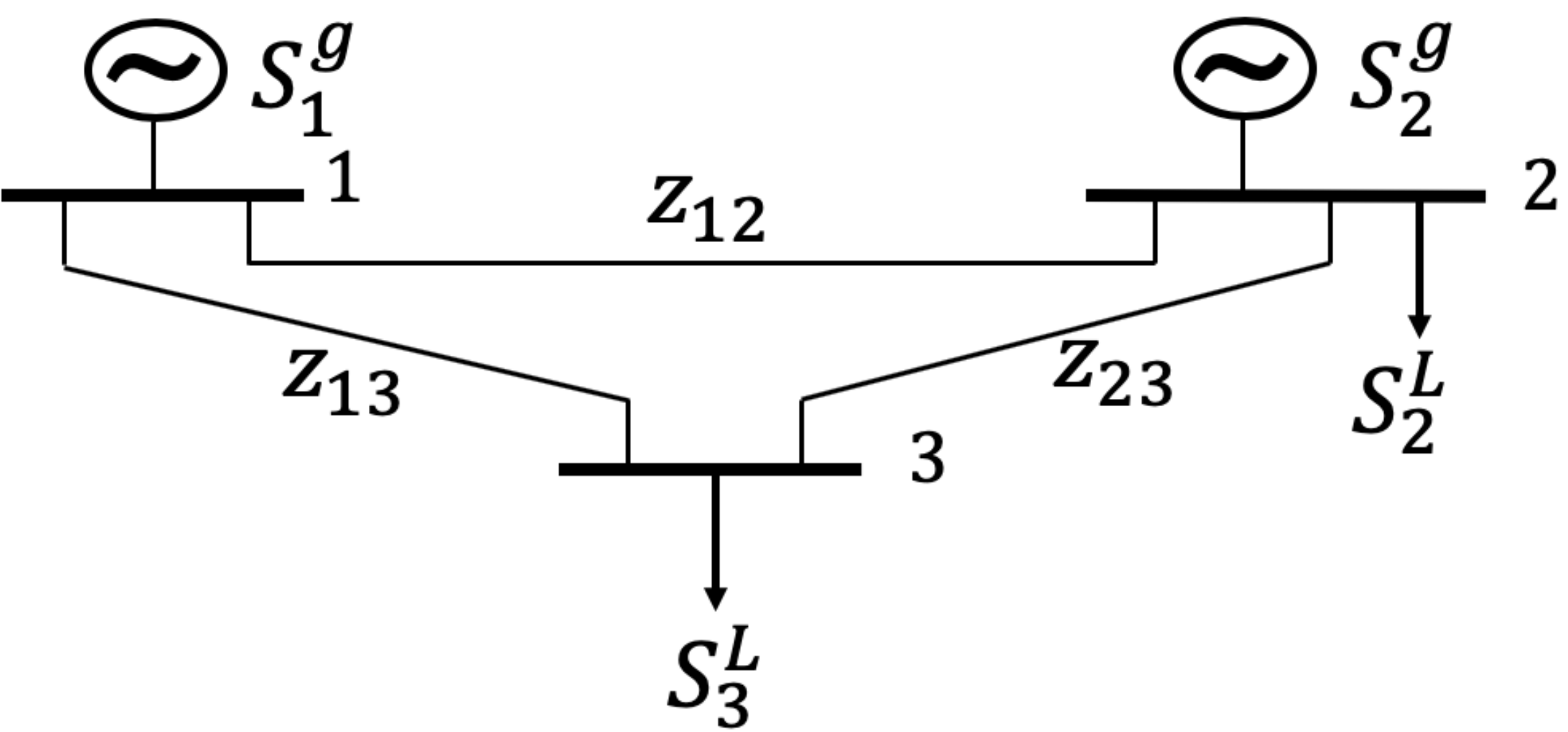}
\caption{A Single diagram of a 3 bus system with $Z_{12}=0.008+j0.024$, $Z_{23}=0.006+j0.018$, and $Z_{13}=0.002+j0.060$.}
\label{fig:3bus}
\end{figure}

\section{Machine Learning (ML)}
Machine Learning methods are conceptually very similar to statistics with some minor differences. However, these differences are essential and distinguishes machine learning from conventional statistics. Machine learning approximates a function within a hypothesis space as opposed to modeling a predefined function in a standard statistical method. This also leads the machine learning methods to value the accuracy of the prediction rather then the quality of the fit which is the primary objective in a statistical approach \cite{rahman,rahman11}. The idea of employing machine learning in power systems dates back to the late 20th century \cite{aips1, aips2}. However, it failed to attract enough interest due to the computation complexities associated with modeling and data handling needed for effectively applying machine learning algorithms. The recent advancements in computational capabilities have led the academia to reconsider machine learning methods for solving power system problems, particularly the power system operations. Since, it is a fairly new topic, many of the problems remain open-ended, particularly on how machine learning can help in achieving a real-time optimal solution to the ACOPF problems. Here we will briefly introduce various ML algorithms that we are going to evaluate its fitting quality for an optimal power flow dataset.

\begin{figure}[h!]
\centering
\includegraphics[scale=0.28]{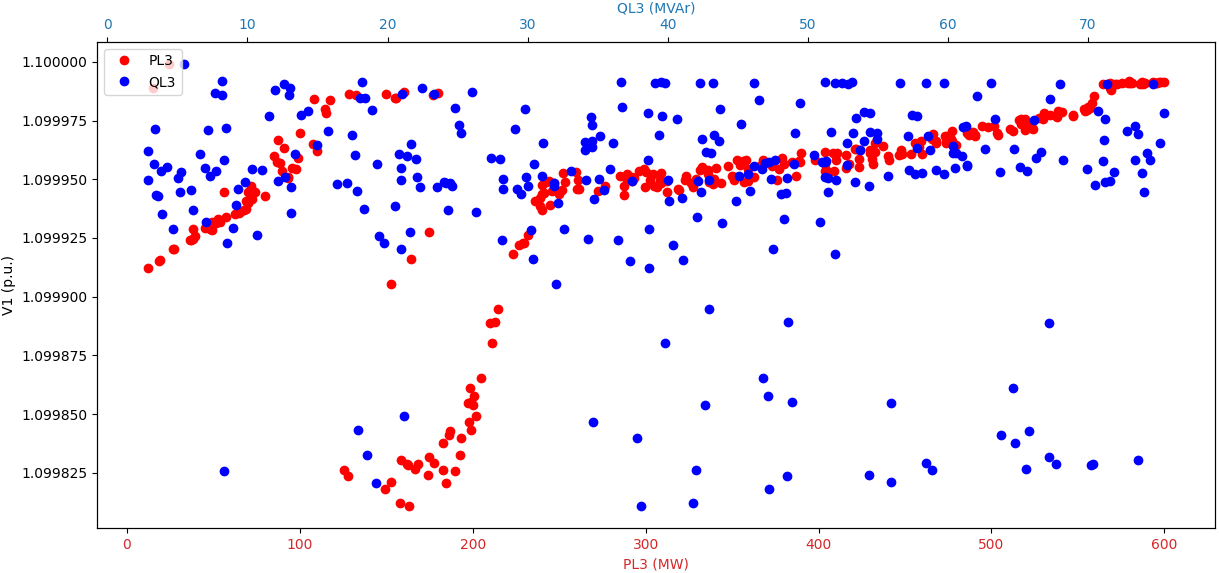}
\caption{A plot of $V_1$ vs $P_3^L$ (Red) and $Q_3^L$ (Blue) in our training data for the non-congested 3 bus system shown in fig. \ref{fig:3bus}}
\label{fig:plot1_1}
\end{figure}

\begin{figure}[h!]
\centering
\includegraphics[scale=0.28]{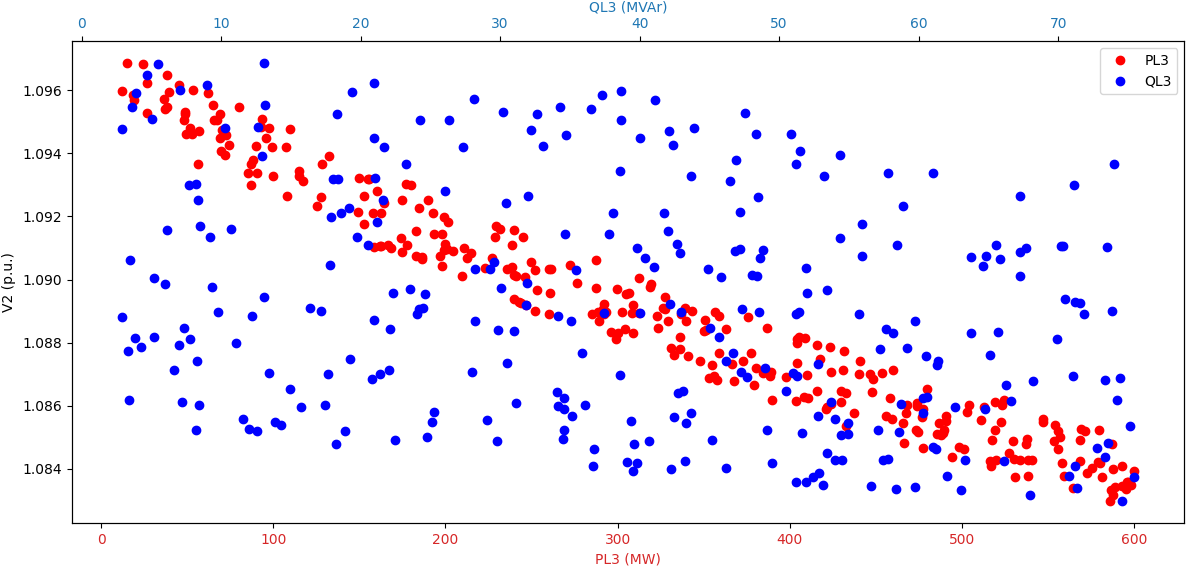}
\caption{A plot of $V_2$ vs $P_3^L$ (Red) and $Q_3^L$ (Blue) in our training data for the non-congested 3 bus system shown in fig. \ref{fig:3bus}}
\label{fig:plot1_2}
\end{figure}

\begin{figure}[h!]
\centering
\includegraphics[scale=0.28]{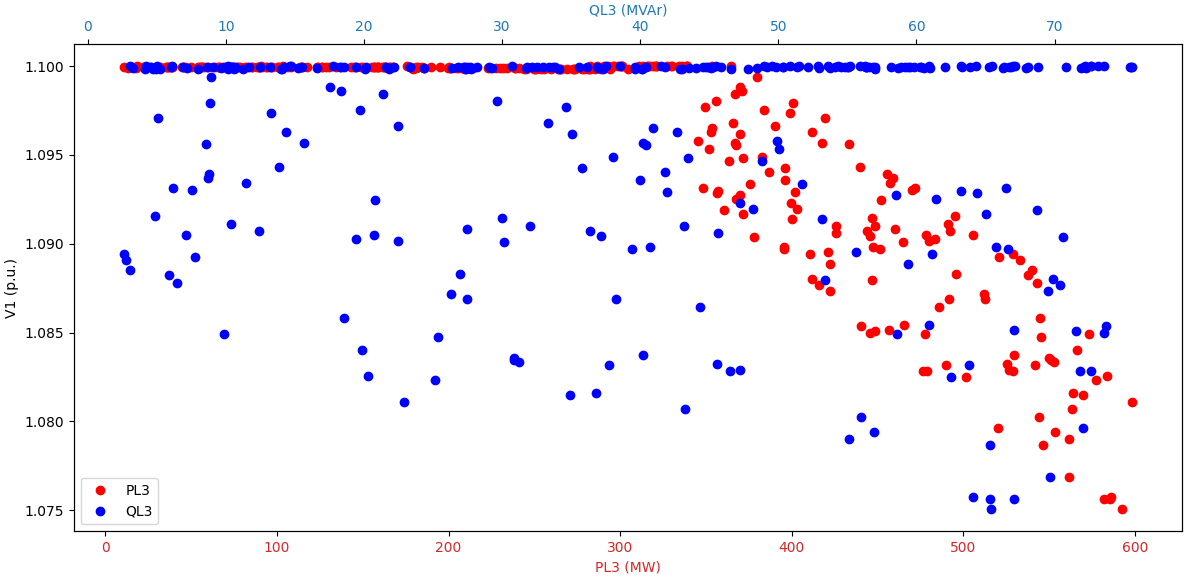}
\caption{A plot of $V_1$ vs $P_3^L$ (Red) and $Q_3^L$ (Blue) in our training data for the congested 3 bus system shown in fig. \ref{fig:3bus}}
\label{fig:plot2_1}
\end{figure}

\begin{figure}[h!]
\centering
\includegraphics[scale=0.28]{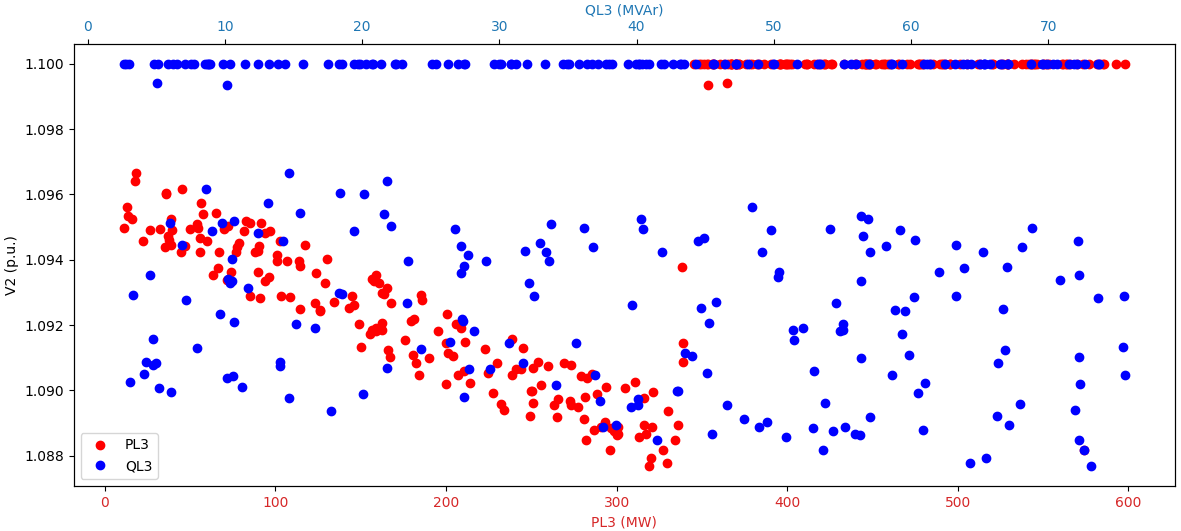}
\caption{A plot of $V_2$ vs $P_3^L$ (Red) and $Q_3^L$ (Blue) in our training data for the congested 3 bus system shown in fig. \ref{fig:3bus}}
\label{fig:plot2_2}
\end{figure}

\begin{figure}[h!]
\centering
\includegraphics[scale=0.28]{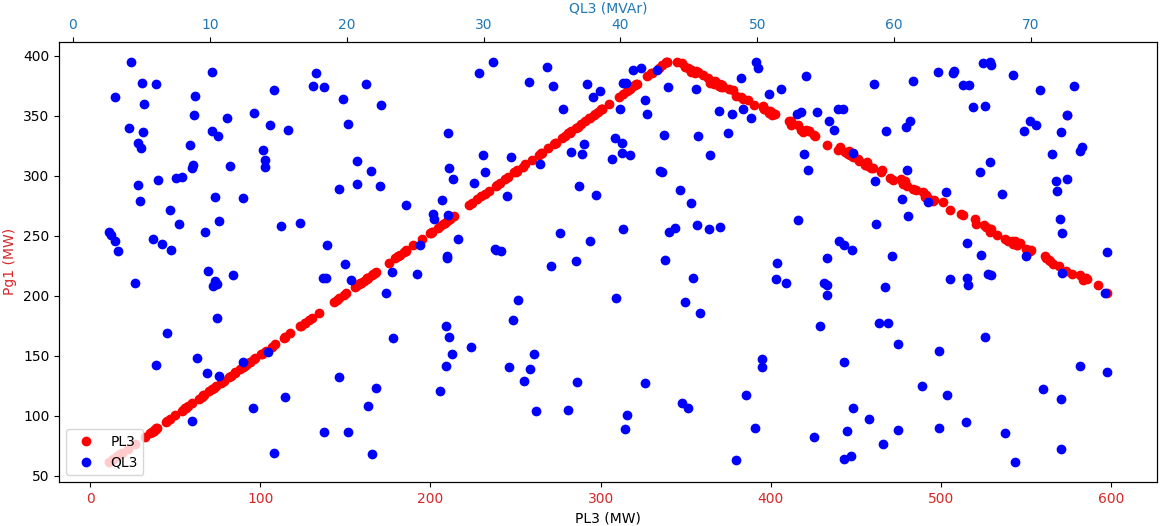}
\caption{A plot of $P^g_1$ vs $P_3^L$ (Red) and $Q_3^L$ (Blue) in our training data for the non-congested 3 bus system shown in fig. \ref{fig:3bus}}
\label{fig:plot2_3}
\end{figure}

\subsection{Linear Regression (LR)}
Linear Regression (LR) is one of the fundamental statistics approach to modeling different relationships and patterns in data. As the name suggests, in Linear Regression, the label is seen as a linear combinations of features. In the Scikit-Learn library the implementation of LR fits a linear model with coefficients $\omega = (\omega_1,...,\omega_p)$ to minimize the residual sum of squares between the observed targets in the dataset, and the targets predicted by the linear approximation. Mathematically it solves a problem of the form \cite{scikit}:
\begin{equation}
\begin{aligned}
& \underset{\omega}{\text{minimize}}
& & ||X\omega-y||_2^2
\end{aligned}
\end{equation}

\subsection{Support Vector Machines (SVM)}
Support vector machines (SVMs) are a set of supervised learning methods used for classification, regression, and outliers detection. It is one of the most popular algorithms in machine learning and it presents one of the most robust prediction methods based on the theories proposed by Vapnik and Chervonekis. 
The advantages of support vector machines are:
\begin{itemize}
\item Effective in high dimensional spaces and memory efficient because it only uses a smaller subset of the total training data  \cite{scikit}.
\item It is one of the methods which remains effective when the number of samples is less than the total number of features \cite{scikit}.
\end{itemize}
The disadvantages of support vector machines include:
\begin{itemize}
\item If the number of features is much greater than the number of samples, over-fitting can be become a challange and needs to be addressed  \cite{scikit}.
\item Calculating probability estimates in SVMs can become very computationally expensive  \cite{scikit}.
\end{itemize}

\subsection{Nearest Neighbors Regression (NNR)}
This method is also called k-Nearest Neighbors method which is one of the non-parametric algorithms and was first proposed by Thomas Cover. The main concept in this approach is to find a predefined number of training samples having the closest distance to the new data point, and predict its target value. The number of samples can be a user-defined constant (k-Nearest Neighbors learning). The distance can, in general, be any metric measure: standard Euclidean distance is the most common choice. Neighbors-based methods are known as non-generalizing machine learning methods, since they simply “remember” all of its training data (possibly transformed into a fast indexing structure such as a Ball Tree or KD Tree) \cite{scikit}.

Even though this method is very simple compared to other algorithms, it has shown impressive success in solving large scale problems.  The method is also effective when less data is available for handling large scale problems \cite{scikit}. 

\subsection{Decision Trees Regression (DTR)}
Decision Trees (DTs) are also a non-parametric the most popular supervised learning method used for classification and regression. The algorithm learns simple rules with the help of data features to predict the value of a target variable. This creates a model which is called a Decision Tree Model~\cite{scikit}.

\subsection{Neural network models (NN)}
Neural networks (NN) are biologically-inspired deep learning architectures where interconnected “neurons” reproduce the learning process using the data in a interconnected set of "parallelly distributed processors" in order to predict real solutions based on this knowledge \cite{rahman, rahman12, rahman13, rahman14}. 

\begin{table*}[h!]
\begin{center}
\caption{The table compares the feasible solution of OPF for a 3 bus non-congested system calculated by Ipopt, Power Flow (PF) calculated by Machine Learning Prediction, and Power Flow calculated by DCOPF solutions.}
\label{table:NC}
\begin{tabular}{|c|c|c|c|c|c|c|c|}
\hline \hline
  \multirow{3}{*}{Sno.} &\multirow{3}*{\begin{tabular}{l} $P^L_3$\\(MW) \end{tabular}}   &\multirow{3}*{\begin{tabular}{l}$Q^L_3$\\(MW) \end{tabular}}  &\multirow{3}*{\begin{tabular}{l} Optimal\\Cost\\(USD) \end{tabular}}  &\multicolumn{2}{c|}{Machine Learning + PF} &\multicolumn{2}{c|}{DCOPF + Power Flow} \\ \cline{5-8}
    &    &    &  &Cost  &Slack Change    &Cost  &Slack Change\\ 
  
    &    &    &     &(USD)  &(MW)    &(USD)  &(MW)  \\ \hline
  
1  &35.91  &377.37  &434.75  &434.75  &0.001  &436.65  &2.169 \\ \hline
2  &53.17  &598.25  &666.23  &666.23  &0.000  &670.83  &3.483 \\ \hline
3  &74.22  &389.68  &447.70  &447.71  &0.000  &449.84  &2.311 \\ \hline
4  &73.99  &56.42  &106.96  &106.96  &0.001  &107.15  &0.686 \\ \hline
5  &54.38  &174.16  &226.10  &226.10  &0.000  &226.65  &1.111 \\ \hline
6  &61.06  &498.01  &560.65  &560.65  &0.001  &563.92  &2.904 \\ \hline
7  &26.32  &65.54  &116.01  &116.01  &0.000  &116.15  &0.530 \\ \hline
8  &61.39  &347.73  &404.17  &404.17  &0.000  &405.88  &2.049 \\ \hline
9  &18.03  &155.18  &206.72  &206.72  &0.000  &207.13  &0.947 \\ \hline
10  &43.34  &486.89  &548.89  &548.89  &0.001  &551.96  &2.807 \\ \hline
11  &4.28  &423.63  &482.69  &482.69  &0.001  &484.95  &2.391 \\ \hline
12  &35.11  &329.01  &384.73  &384.73  &0.000  &386.21  &1.899 \\ \hline
13  &6.64  &459.83  &520.44  &520.45  &0.001  &523.09  &2.600 \\ \hline
14  &13.05  &14.83  &64.99  &64.99  &0.000  &65.04  &0.319 \\ \hline
15  &32.20  &70.30  &120.82  &120.82  &0.000  &120.98  &0.564 \\ \hline
 \hline
\end{tabular}
\end{center}
\end{table*}

\begin{table*}[h!]
\begin{center}
\caption{The table compares the feasible solution of OPF for a congested 3 bus system calculated by Ipopt, Power Flow (PF) calculated by Machine Learning Prediction, and Power Flow calculated by DCOPF solutions.}
\label{table:C}
\begin{tabular}{|c|c|c|c|c|c|c|c|c|c|}
\hline \hline
  \multirow{3}{*}{Sno.} &\multirow{3}*{\begin{tabular}{l} $P^L_3$\\(MW) \end{tabular}}   &\multirow{3}*{\begin{tabular}{l}$Q^L_3$\\(MW) \end{tabular}}  &\multirow{3}*{\begin{tabular}{l} Optimal\\Cost\\(USD) \end{tabular}}  &\multicolumn{3}{c|}{Machine Learning + Power FLow} &\multicolumn{3}{c|}{DCOPF + Power Flow} \\ \cline{5-10}
    &    &    &  &Cost  &Slack Change    &Violation    &Cost  &Slack Change   &Violation\\ 
  
    &    &    &     &(USD)  &(MW)    & (MW)    &(USD)  &(MW)   &(MW)\\ \hline
  
1  &63.24  &103.33  &154.27  &149.19  &5.133  &0.000  &185.09  &15.879  &0.000 \\ \hline
2  &59.35  &155.43  &207.07  &205.65  &1.440  &0.000  &238.93  &16.750  &0.000 \\ \hline
3  &19.84  &566.04  &1025.73  &1034.02  &8.200  &0.000  &1027.95  &25.673  &37.273 \\ \hline
4  &29.06  &106.64  &157.52  &163.28  &5.813  &0.000  &188.36  &15.860  &0.000 \\ \hline
5  &2.78  &533.02  &933.10  &929.93  &3.165  &0.781  &935.66  &24.930  &41.499 \\ \hline
6  &18.52  &163.42  &215.09  &215.77  &0.692  &0.000  &246.90  &16.737  &0.000 \\ \hline
7  &64.09  &585.99  &1081.20  &1080.62  &0.587  &0.140  &1085.10  &26.813  &30.385 \\ \hline
8  &68.06  &35.32  &85.70  &82.14  &3.586  &0.000  &115.77  &15.224  &0.000 \\ \hline
9  &9.67  &200.72  &253.06  &248.21  &4.915  &0.000  &285.54  &17.410  &0.000 \\ \hline
10  &52.95  &210.28  &262.92  &265.91  &3.032  &0.000  &295.94  &17.825  &0.000 \\ \hline
11  &29.29  &241.07  &294.32  &291.57  &2.803  &0.000  &328.18  &18.553  &3.985 \\ \hline
12  &36.16  &54.39  &104.80  &105.32  &0.524  &0.000  &134.90  &15.253  &0.000 \\ \hline
13  &22.43  &479.85  &783.10  &781.79  &1.302  &0.326  &789.31  &24.865  &41.802 \\ \hline
14  &56.33  &93.74  &144.55  &141.88  &2.688  &0.000  &175.28  &15.768  &0.000 \\ \hline
15  &38.01  &382.37  &511.56  &514.53  &2.978  &0.000  &523.36  &25.924  &45.108 \\ \hline
 \hline

\end{tabular}
\end{center}
\end{table*}

\section{Results}
For the simplicity of our analysis, we use a 3 bus system in our evaluation shown in Fig. \ref{fig:3bus}.

To generate data for our training we have run about 600 optimization solutions, each with a random load of $S_3^L$; 300 for non-congested 3 bus system and 300 for congested 3 bus system. In the congested scenario we have limited the flow of power in branch1-3 to 160 MVA. The plots shown in Figs. \ref{fig:plot1_1}, \ref{fig:plot1_2}, \ref{fig:plot2_1}, \ref{fig:plot2_2}, and \ref{fig:plot2_3} presents the relationship between features and targets in our training data.

\begin{table}[h!]
\begin{center}
\caption{The table shows the calculated accuracy of each algorithm in predicting different variables in both non-congested and congested scenarios for the 3 bus system shown in Fig.\ref{fig:3bus}.}
\label{table:comp:algorithms}
\begin{tabular}{|c|c|c|c|c|c|}
\hline \hline
 \multirow{2}{*}{Algorithms} &\multicolumn{2}{c|}{Non-congested} &\multicolumn{3}{c|}{Congested} \\ \cline{2-6}
    &$V_1$  &$V_2$    &$V_1$  &$V_2$ &$P_g^1$\\ \hline
  
LR  &99.9976 &99.9733 &99.7475 &99.7579 &68.9835  \\ \hline
SVM  &99.9945 &99.6744 &99.1781 &99.6253 &97.0389 \\ \hline
NNR  &99.9996 &99.9884 &99.9791 &99.9861 &98.6489 \\ \hline
DTR  &99.9985 &99.9433 &99.8793 &99.9650 &93.6066 \\ \hline
NN  &83.3745 &95.6851 &94.1331 &96.5417 &76.9921 \\
\hline
 \hline
\end{tabular}
\end{center}
\end{table}

\begin{figure}[h!]
\centering
\includegraphics[scale=0.28]{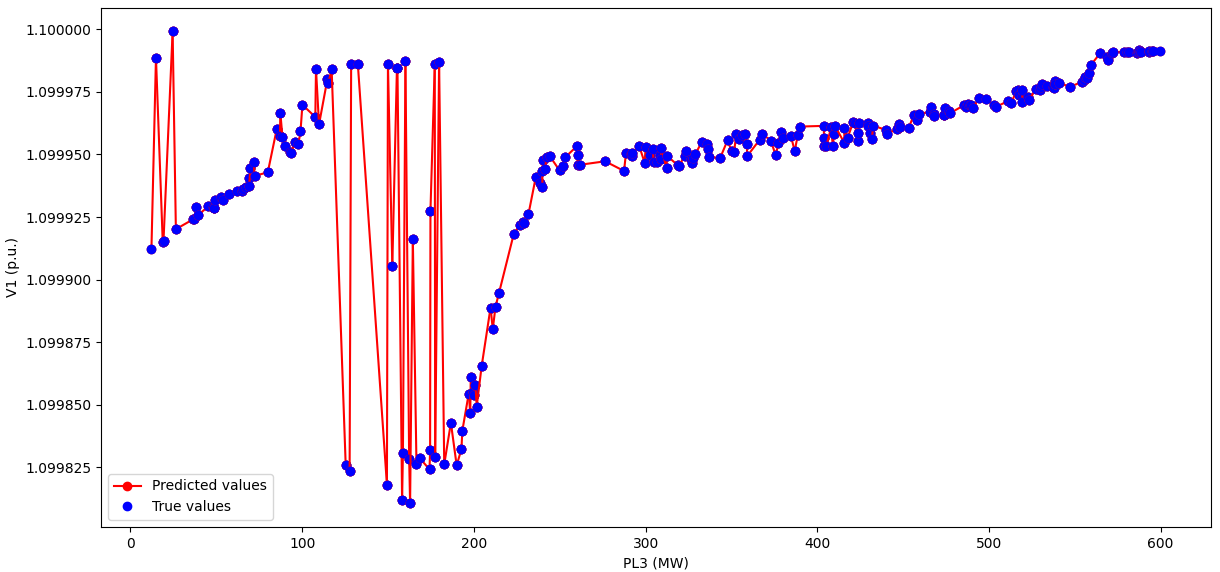}
\caption{A plot showing the quality of fitting for $P_3^L$ vs $V_1$ in the non-congested 3 bus system by the NNR method.}
\label{fig:plot3_1}
\end{figure}

\begin{figure}[h!]
\centering
\includegraphics[scale=0.28]{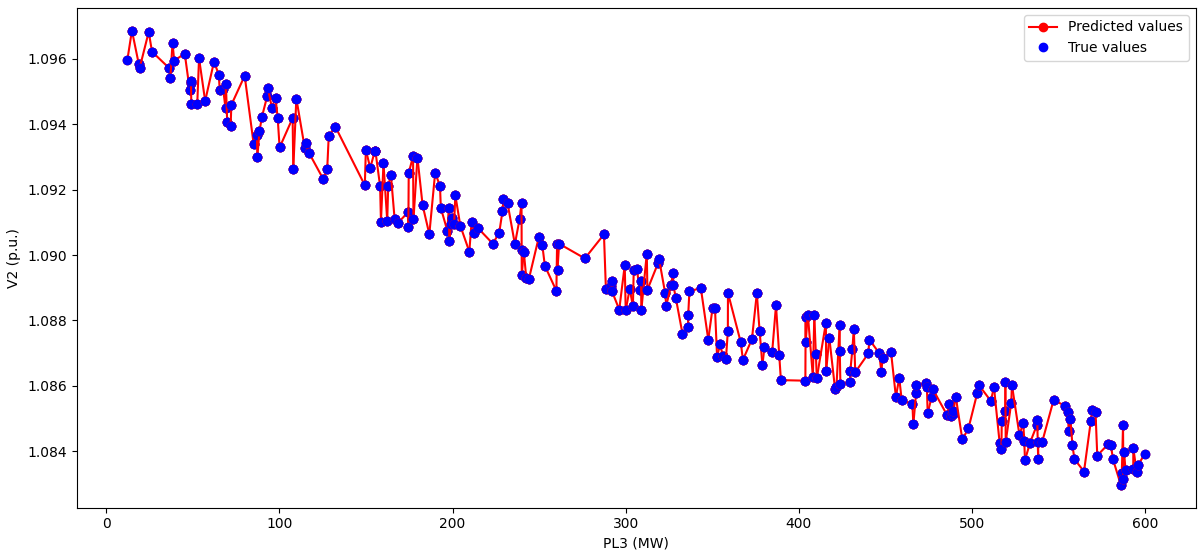}
\caption{A plot showing the quality of fitting for $P_3^L$ vs $V_2$ in the non-congested 3 bus system by the NNR method.}
\label{fig:plot3_2}
\end{figure}

\begin{figure}[h!]
\centering
\includegraphics[scale=0.28]{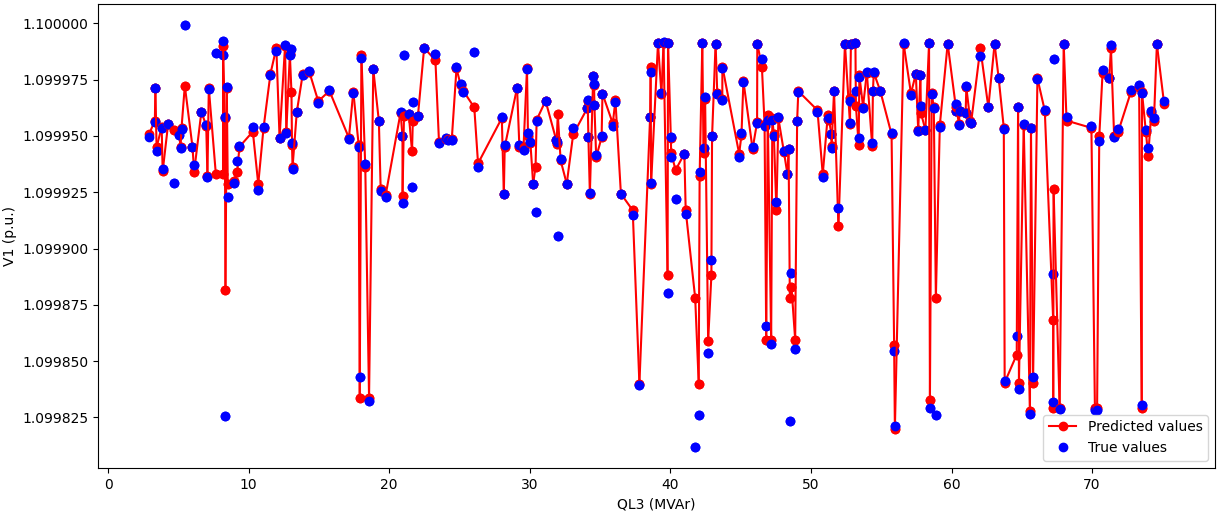}
\caption{A plot showing the quality of fitting for $Q_3^L$ vs $V_1$ in the non-congested 3 bus system by the NNR method.}
\label{fig:plot3_3}
\end{figure}

\begin{figure}[h!]
\centering
\includegraphics[scale=0.28]{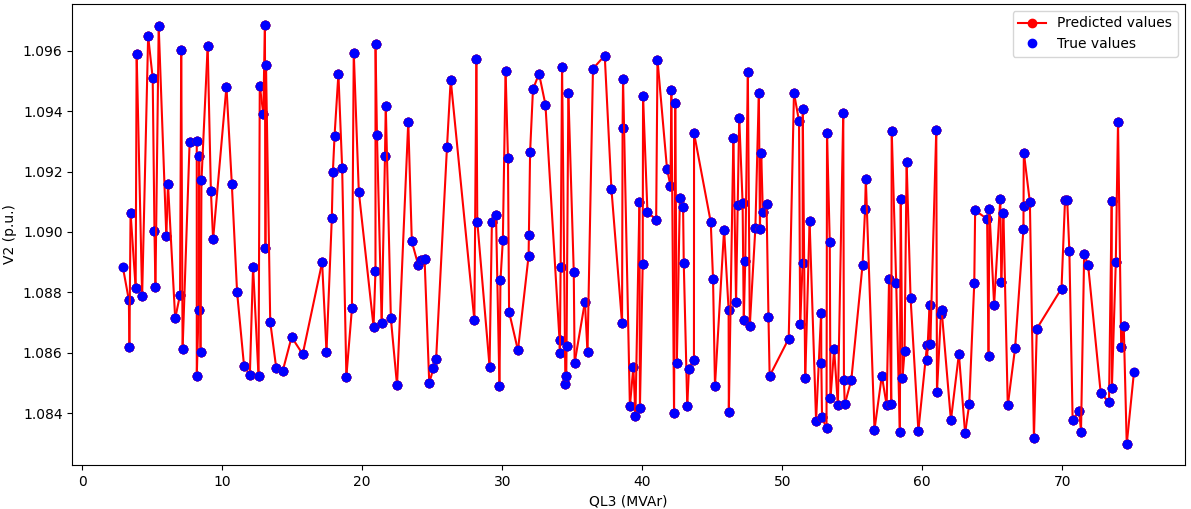}
\caption{A plot showing the quality of fitting for $Q_3^L$ vs $V_2$ in the non-congested 3 bus system by the NNR method.}
\label{fig:plot3_4}
\end{figure}

We first evaluated the quality of the fit by different popular algorithms available in Scikit Library. The comparison is shown in Table \ref{table:comp:algorithms}. To fine-tune the parameters of each estimator we have applied the exhaustive grid search method. The grid search method exhaustively generates combinations from a grid of parameter values to fine-tune the hyper-parameters of an estimator.

By looking at the results presented in Table \ref{table:comp:algorithms} we can observe that the Nearest Neighbors Regression (NNR) method gives the highest accuracy of prediction. NNR being the best performance estimator, we now evaluate the results of this algorithm in solving OPF and benchmark it with DCOPF solutions for a 3 bus system.  
Table \ref{table:NC} compares the dispatch cost calculated by Ipopt, ML (our proposed method) and using DCOPF solution. To obtain feasibility both ML and DCOPF solutions have been solved by power flow. Slack change shows the feasibility of predicted solution compared with the DCOPF solution. Similarly, Table \ref{table:C}, compares the results for a congested 3 bus system. It has additional columns which calculates the violations of branch 1-3 thermal limit. 

\begin{figure}[h!]
\centering
\includegraphics[scale=0.28]{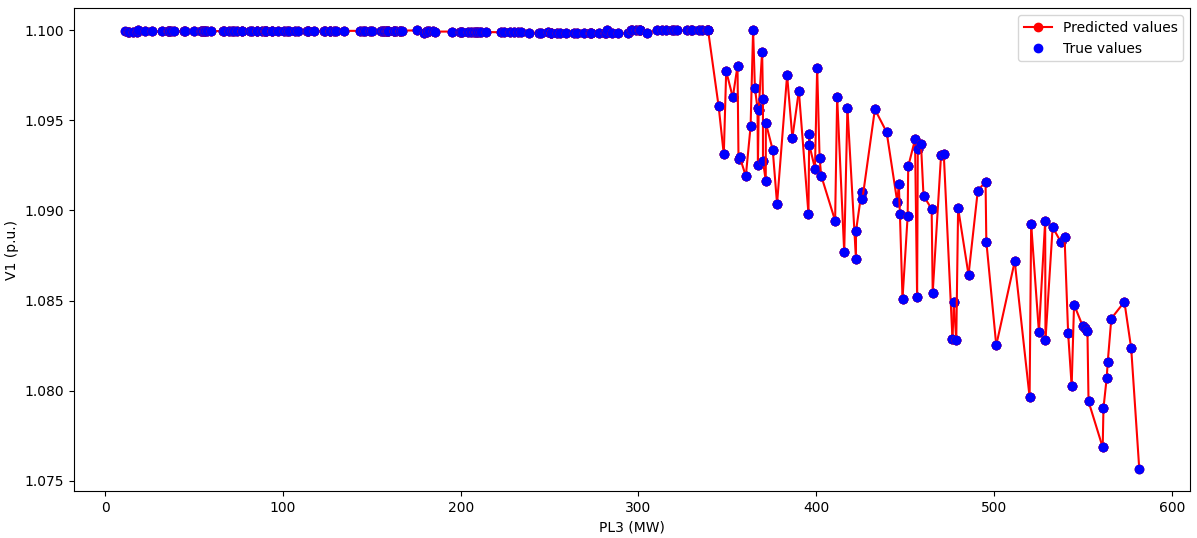}
\caption{A plot showing the quality of fitting for $P_3^L$ vs $V_1$ in the congested 3 bus system by the NNR method.}
\label{fig:plot4_1}
\end{figure}

\begin{figure}[h!]
\centering
\includegraphics[scale=0.28]{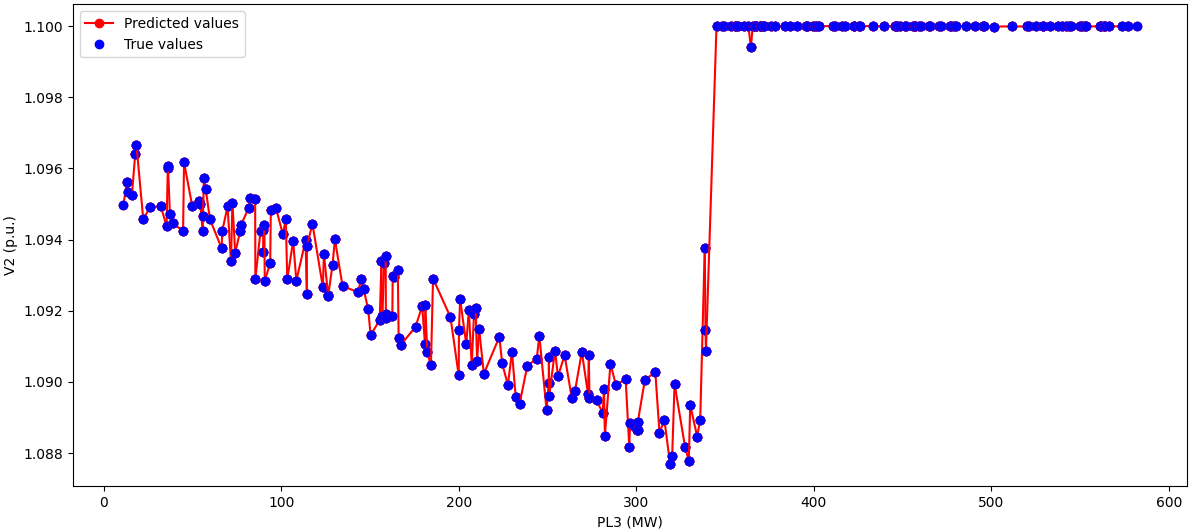}
\caption{A plot showing the quality of fitting for $P_3^L$ vs $V_2$ in the congested 3 bus system by the NNR method.}
\label{fig:plot4_2}
\end{figure}

\begin{figure}[h!]
\centering
\includegraphics[scale=0.28]{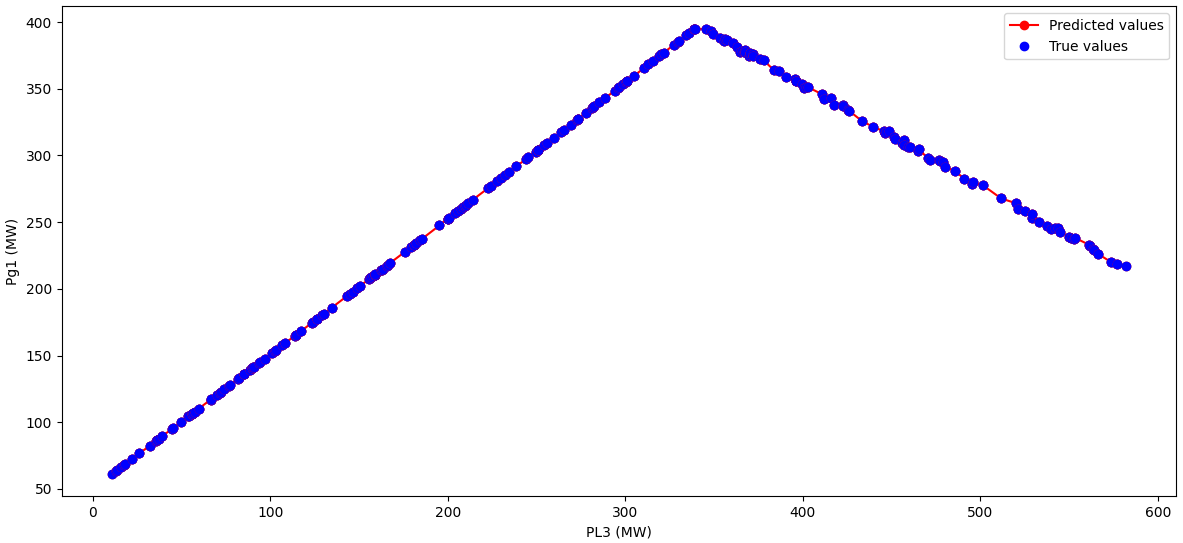}
\caption{A plot showing the quality of fitting for $P_3^L$ vs $P^g_1$ in the congested 3 bus system by the NNR method.}
\label{fig:plot4_3}
\end{figure}

\begin{figure}[h!]
\centering
\includegraphics[scale=0.28]{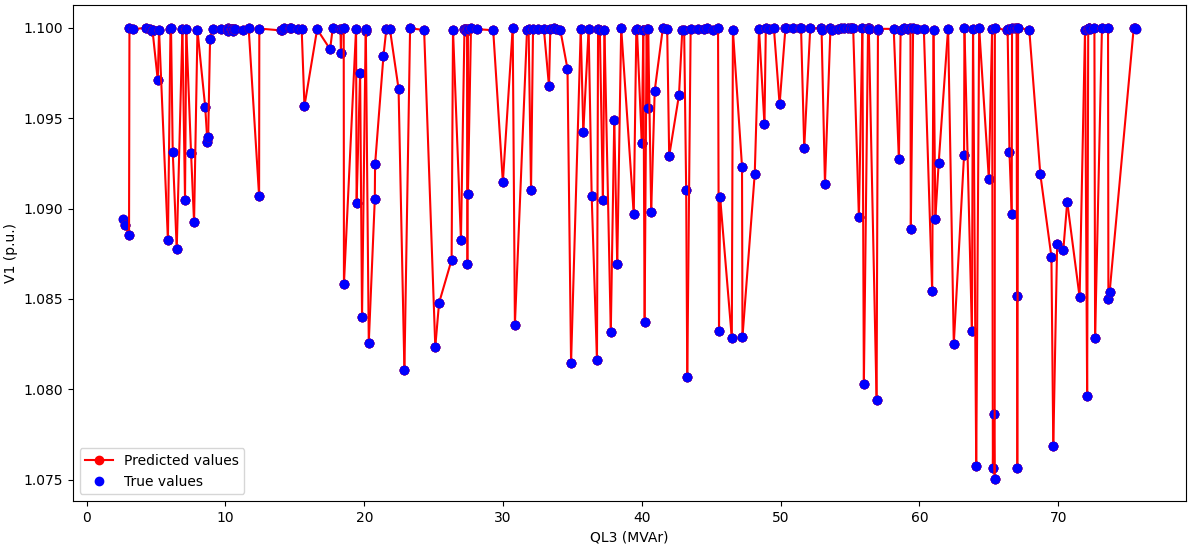}
\caption{A plot showing the quality of fitting for $Q_3^L$ vs $V_1$ in the congested 3 bus system by the NNR method.}
\label{fig:plot4_4}
\end{figure}

\begin{figure}[h!]
\centering
\includegraphics[scale=0.28]{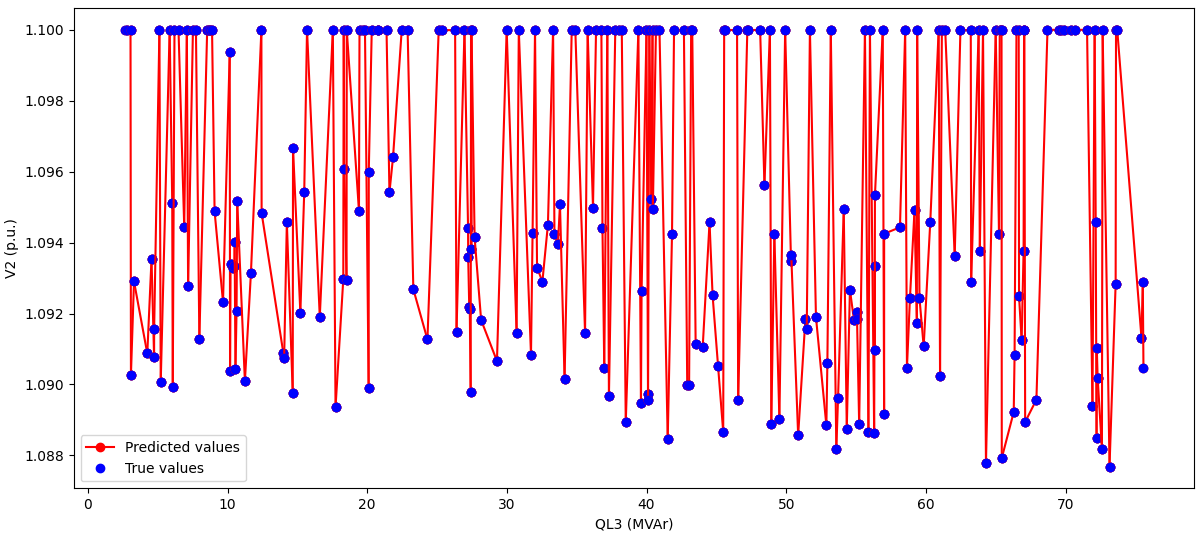}
\caption{A plot showing the quality of fitting for $Q_3^L$ vs $V_2$ in the congested 3 bus system by the NNR method.}
\label{fig:plot4_5}
\end{figure}

\begin{figure}[h!]
\centering
\includegraphics[scale=0.28]{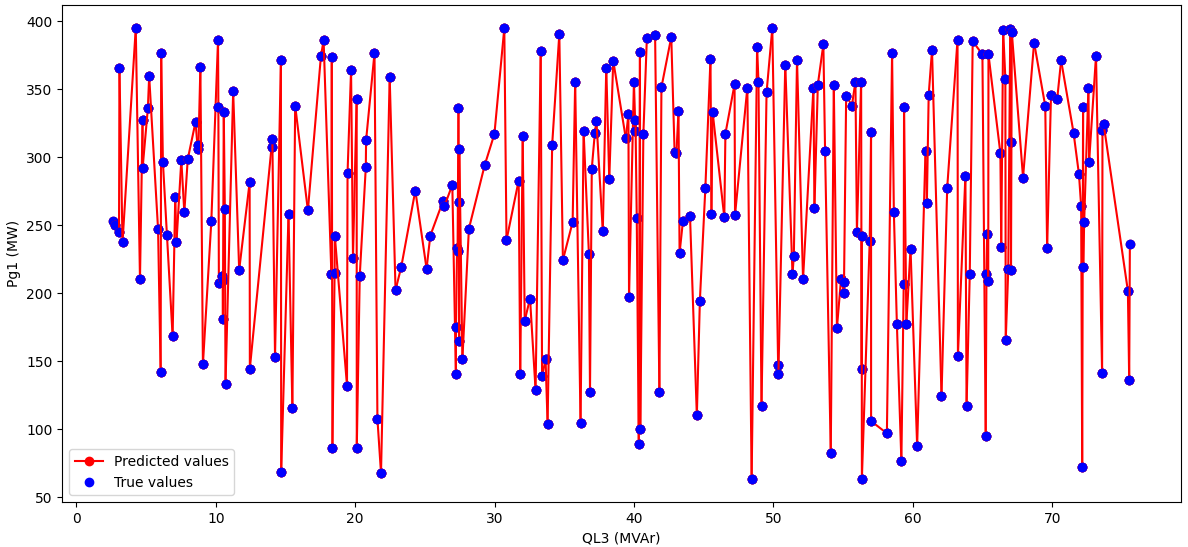}
\caption{A plot showing the quality of fitting for $Q_3^L$ vs $P^g_1$ in the congested 3 bus system by the NNR method.}
\label{fig:plot4_6}
\end{figure}

Figs. \ref{fig:plot3_1}, \ref{fig:plot3_2}, \ref{fig:plot3_3}, \ref{fig:plot3_4}, and \ref{fig:plot4_1} show how the estimator's predictions fit the actual data for a non-congested 3 bus system. Similarly, Figs. \ref{fig:plot4_1}, \ref{fig:plot4_2}, \ref{fig:plot4_3}, \ref{fig:plot4_4}, \ref{fig:plot4_5}, and \ref{fig:plot4_6} show how the estimator fits the training data for a congested 3 bus system.

\section{Discussion}

As we can see, the Nearest Neighbors method gives promising results which can be used as initialization for an SLP solver. The method is simple, fast, and provides the highest accuracy than other popular algorithms. The results show a promising approach for the initialization of an SLP solver for an ACOPF problem. The SLP solver can be significantly impacted by the initialization, and as shown in Tables \ref{table:NC} and \ref{table:C}, the predicted results are very close to the optimal solution, therefore, offering a superior method for initializing it. Additionally, they have a small feasibility gap (i.e., measured by the changes in the slack) and minimal violation of thermal limits.
The choice of algorithm is important, because different algorithms are chosen depending on their performance and the quality of their fit for various applications, which can be impacted by a few factors such as the availability of training data, noise in the data, etc. The purpose of showing the score (accuracy) of each algorithm for the 3 bus system in Table \ref{table:comp:algorithms} is to show the quality of their fit for the OPF solution of a 3 bus system. Should we decide in the future to choose a different algorithm because of the availability of the data for training or any other reason for larger systems, this table, in our opinion, may give us an insight into the trade-off in the performance (quality of the fit) that we are looking into versus the choice of algorithm. The approach can also be used in the industry in real-time directly in combination with some heuristic approaches to make the solution feasible. As can be seen, the results are far better than the DCOPF solutions used in the industry. In case sufficient data is not available, we can use other methods or an SLP-ACOPF with DCOPF initialization in an offline mode to generate data for our proposed method. Once we have sufficient data to train our estimator, we use it as an initialization for the SLP method to enhance its performance for the given network. In our data which is generated by the direct solution of an optimization model, we expect to have zero to minimal noise. Leveraging this advantage we can even use Generative Adversarial Networks (GAN) to generate data from a small sample solutions as a worst case scenario for addressing the lack of sufficient data for training.

\section{Conclusion}
We have evaluated the performance of different machine learning estimators for the solution of an ACOPF problem. In our evaluation, we have fine-tuned the hyper-parameters of each algorithm in the Scikit-Learn Library. We found that the Nearest Neighbors Regression method gives the highest accuracy of prediction for the 3 bus system. The results show that the proposed method can provide an initialization far better than the DCOPF approach for initializing an SLP solver for solving ACOPF problems. The prediction time of these estimators are in microseconds which makes it superior to the DCOPF algorithms used in the industry.



\ifCLASSOPTIONcaptionsoff
  \newpage
\fi

\bibliographystyle{IEEEtran}
\bibliography{IEEEabrv,main}

\end{document}